\documentclass[floats,floatfix,showpacs,amssymb,prd,superscriptaddress,twocolumn,aps]{revtex4-1}

\usepackage{graphicx}  
\usepackage{dcolumn}   
\usepackage{bm}        
\usepackage{amssymb}   

\usepackage{lipsum}

\usepackage{epsf}

\usepackage{graphicx}

\usepackage{amsmath}

\usepackage[linktocpage]{hyperref}
\usepackage[caption=false]{subfig}
\usepackage[usenames]{color}

\newcommand{\insertplot}[5]{\begin{figure}
 \hfill\hbox to 0.05in{\vbox to #5in{\vfill
 \inputplot{#1}{#4}{#5}}\hfill}
 \hfill\vspace{-.1in}
 \caption{#2}\label{#3}
 \end{figure}}
 \newcommand{\inputplot}[3]{
 \special{ps: plotfile #1}
}
\newcounter{fig}

\usepackage{graphicx}

\begin{document} 
\title{
New black holes in $D=5$ minimal gauged supergravity: 
\\
deformed boundaries and 'frozen' horizons
}
 \vspace{1.5truecm}

\author{Jose Luis Bl\'azquez-Salcedo} 
 \affiliation{Institut f\"ur  Physik, Universit\"at Oldenburg  Postfach 2503,
D-26111 Oldenburg, Germany}

\author{Jutta Kunz} 
 \affiliation{Institut f\"ur  Physik, Universit\"at Oldenburg  Postfach 2503, 
 D-26111 Oldenburg, Germany}

\author{Francisco Navarro-L\'erida} 
 \affiliation{Departamento de F\'{\i}sica Te\'orica I and UPARCOS, Ciencias F\'{\i}sicas, \\
 Universidad Complutense de Madrid, E-28040 Madrid, Spain}

\author{Eugen Radu} 
 \affiliation{Departamento de F\'isica da Universidade de Aveiro and CIDMA, \\
 Campus de Santiago, 3810-183 Aveiro, Portugal}

\date{\today}


\begin{abstract} 
A new class of black hole solutions of the five dimensional minimal gauged supergravity is presented. 
They are 
 characterized by the mass, the electric charge, two equal magnitude angular momenta
and the magnitude of the magnetic potential at infinity.
These black holes possess a
horizon of spherical topology;
however, both the horizon and
 the sphere at infinity  can be arbitrarily squashed,
with nonextremal solutions
interpolating between black strings and black branes.
A particular set of extremal configurations corresponds to a
new one-parameter family of supersymmetric black holes.
While their conserved charges are determined by the squashing 
of the sphere at infinity,
these supersymmetric solutions possess the same horizon geometry.

\end{abstract}

\maketitle

  \medskip 
{\bf Introduction.--}
%
There has recently been considerable interest in 
solutions of the five-dimensional gauged supergravity models,
mainly motivated by the AdS/CFT correspondence
\cite{Maldacena:1997re},
\cite{Witten:1998qj}.
The black holes (BHs) play  a central role in this context,
providing the thermodynamic saddle
points of the dual four-dimensional theory. 

The Schwarzschild-AdS BH provides the  simplest example,
possessing a spherical horizon and a single global charge: the mass. 
As expected, the inclusion of other charges
($e.g.$ the angular momenta)
generically deforms the horizon shape.
Despite the existence of a number of partial results,
the study of the horizon geometrical properties 
(in particular its deformation)
in conjunction with other BH properties
is a rather poorly explored subject, 
presumably
due to the complexity of the problem.
However, this study is greatly simplified 
by restricting to  
BHs with a spherical horizon topology which possess 
two equal-magnitude angular momenta.
Then one can use 
 a cohomogeneity-1 Ansatz which
factorizes the angular dependence of the metric and the gauge potential
and leads to
a homogeneous squashing of the horizon geometry.
Then, without any loss of generality,
the induced horizon metric can be written as
\begin{eqnarray} 
\label{horizon-metric}
ds^2_{{\cal H}}=\frac{L_H^2}{4}
\left(
 \sigma_1^2+\sigma_2^2 +\epsilon_H^2 \sigma_3^2
\right),
\end{eqnarray}
with $L_H>0$ and the left invariant one-forms  
$
\sigma_1=\cos \psi d  \theta+\sin\psi \sin   \theta d \phi,
$
$
\sigma_2=-\sin \psi d \theta+\cos\psi \sin   \theta d \phi,
$
$
\sigma_3=d\psi  + \cos \theta d \phi;
$
(where coordinates $ \theta$, $\phi$, $\psi$ are the Euler angles on $S^3$,
with the usual range).
The  deformation parameter $\epsilon_H$ 
 gives the ratio of the $S^1$ and the round $S^2$ parts of the (squashed $S^3$-) horizon metric, 
while  the horizon area is $A_H=2\pi^2 L_H^3 \epsilon_H$.

A black hole in minimal gauged supergravity  
with the horizon geometry (\ref{horizon-metric})
	has been constructed in closed form  
	by Cveti\v c, L\"u and Pope in Ref.  \cite{Cvetic:2004hs}.
This solution is characterized by three non-trivial parameters, namely the
mass $M$, the electric charge $Q$, and a rotation parameter $J$. 
An extension which
possesses
 an extra-parameter $\Phi_m$ associated with a non-zero magnitude
of the magnetic potential at infinity
has been reported in 
the recent work \cite{Blazquez-Salcedo:2017cqm}.

These solutions  
possess a conformal boundary geometry which is the
 static Einstein universe.
However, a remarkable property of the AdS/CFT correspondence 
is that it
does not  constrain the way of approaching
the boundary of spacetime, 
asymptotically $locally$ AdS  (AlAdS) solutions being also relevant.
An interesting case here corresponds to configurations whose
conformal
boundary metric
is the product of time and  
 a squashed sphere 
\begin{eqnarray} 
\label{boundary-sphere}
ds^2_{{\cal B}}=\frac{L^2}{4}
\left(
\sigma_1^2+\sigma_2^2+\epsilon_B^2 \sigma_3^2
\right),
\end{eqnarray}
with $\epsilon_B>0$ a squashing parameter and $L$ the AdS length scale.

The main purpose of this work is to investigate 
the correlation between
the squashing parameters $\epsilon_B$ and $\epsilon_H$
and, more general,
how the BH properties are affected by
the deformation of the boundary sphere.
A new class of BH solutions  
is reported in this context.
Possessing arbitrary values of the squashing parameters
$\epsilon_H$
and
$\epsilon_B$, 
the generic solutions interpolate between black strings and black branes.
 A particular limit describes a new one-parameter family
 of supersymmetric (SUSY) BHs, which possess special properties.

{\bf General solutions.--}
In the minimal case, the bosonic sector  of the $d=5$  gauged supergravity consists of the graviton
and an abelian vector only, with action 
\begin{eqnarray} 
\label{EMCSac}
I= \frac{1}{16\pi } \int_{{\cal M}} d^5x 
\sqrt{-g}\biggl[R +\frac{12}{L^2} 
-F_{\mu \nu} F^{\mu \nu}  
\\
\nonumber
-
\frac{2 }{3\sqrt{3}}\varepsilon^{\mu\nu\alpha\beta\gamma}A_{\mu}F_{\nu\alpha}F_{\beta\gamma} 
 \biggr ],
\end{eqnarray}
where $R$ is the curvature scalar,  
$A $ is the gauge potential and $F=dA$ is the field strength tensor. 

The appropriate Ansatz for the metric and the gauge potential is given by
\cite{Blazquez-Salcedo:2017cqm},
\cite{Blazquez-Salcedo:2016rkj}
\begin{eqnarray}
\label{metrici}
&&ds^2 = F_1(r)dr^2
  + \frac{1}{4}F_2(r)(\sigma_1^2+\sigma_2^2)
 \\
 \nonumber
	&&{~~~~~~~~}
	+\frac{1}{4}F_3(r) \big(\sigma_3-2W(r) dt \big)^2-F_0(r) dt^2,
\\ 
\nonumber
&&
{~~}A=a_0(r)dt +a_k(r) \frac{1}{2} \sigma_3.
\end{eqnarray}
The BHs in 
\cite{Cvetic:2004hs},
\cite{Blazquez-Salcedo:2017cqm}
can be written in this form and have $\epsilon_B=1$,
while  $\epsilon_H$ 
presents a complicated dependence on the global parameters 
($e.g.$ with $\epsilon_H>1$ for $Q=\Phi_m=0$).

We have found
that
 these BHs 
 possess a generalization with a squashed Einstein universe in the boundary metric. 
As such, apart from  $\{M,J,Q,\Phi_m \}$
the new BHs have an additional free geometric parameter, 
the boundary
squashing $\epsilon_B$.
 These asymptotics are compatible with the following approximate 
expression of the metric functions at infinity \cite{gauge}
(which fixes the boundary conditions imposed in the numerics): 
\begin{eqnarray}
&& F_0 \sim \left(\frac{r}{L}\right)^2+\frac{1}{9}(13-4\epsilon_B^2)+\dots{~}, \nonumber \\
&& F_1^{-1}\sim  \left(\frac{L}{r}\right)^2-\frac{1}{9}(14-5\epsilon_B^2) \left(\frac{L}{r}\right)^4+\dots{~}, \nonumber \\
&& F_2 \sim r^2-\frac{5L^2}{4}(1-\epsilon_B^2)+\dots{~}, \nonumber \\
&& F_3  \sim \epsilon_B^2r^2 +\frac{13L^2\epsilon_B^2}{9  }(1-\epsilon_B^2)+\dots{~}, \nonumber \\
&& W \sim  -\frac{\hat j}{r^4} +\dots{~},
\end{eqnarray}
while the U(1) potential behaves as
\begin{eqnarray}
&& a_k \sim -2\Phi_m+ 
\left[\mu-4\Phi_m L^2 \epsilon_B^2 \log \left(\frac{L}{r}\right)\right]\frac{1}{r^2}+\dots{~}, \nonumber \\
&& a_0 \sim V_0+q/r^2+\dots{~}.
\end{eqnarray}
The asymptotically flat solutions necessarily have $\Phi_m=0$;
however, an A(l)AdS spacetime effectively acts like a 'box'
	\cite{Herdeiro:2015vaa},
	\cite{Costa:2015gol},
	\cite{Blazquez-Salcedo:2016vwa},
	\cite{Herdeiro:2016plq}.
	This 
allows for the existence of a nonvanishing asymptotic magnetic field,
$F_{\theta \phi} \to \Phi_m \sin\theta $, such that the parameter $\Phi_m$
can be identified with the magnetic flux at infinity
through the base space $S^2$ of the $S^1$ fibration \cite{Blazquez-Salcedo:2017cqm},
\begin{eqnarray}
\label{flux} 
\Phi_m=\frac{1}{4\pi}  \int_{S^2_\infty} F.
\end{eqnarray} 

The global charges of the solutions 
are encoded in a set of free coefficients which enter their asymptotic expansion, 
being computed by using
the standard holographic renormalisation procedure
\cite{Henningson:1998gx},
\cite{Balasubramanian:1999re},
\cite{Sahoo:2010sp}.  
One finds $e.g.$ the angular momentum and the (holographic) electric charge
\begin{eqnarray}
J=\frac{ \pi  \hat j \epsilon_B^3}{4}~,~~
Q =-\pi \left(q \epsilon_B+\frac{16 }{3\sqrt{3}}\Phi_m^2 \right).
\end{eqnarray}

The solutions possess a horizon located at $r=r_H>0$, 
where, restricting to the nonextremal case,
 $F_0=f_0(r-r_H)^2+\dots$,
while the remaining functions 
are nonzero.
The Hawking temperature is 
$T_H=\frac{1}{2\pi}\sqrt{f_0/F_1(r_H)}$,
while
the horizon metric is
given by
(\ref{horizon-metric}),
with $L_H^2=F_2(r_H)$ and  $\epsilon_H^2=F_3(r_H)/F_2(r_H)$.

\begin{figure}[t!]
\begin{center}
\mbox{\hspace*{-.8cm}
\includegraphics[height=.25\textheight, angle =0]{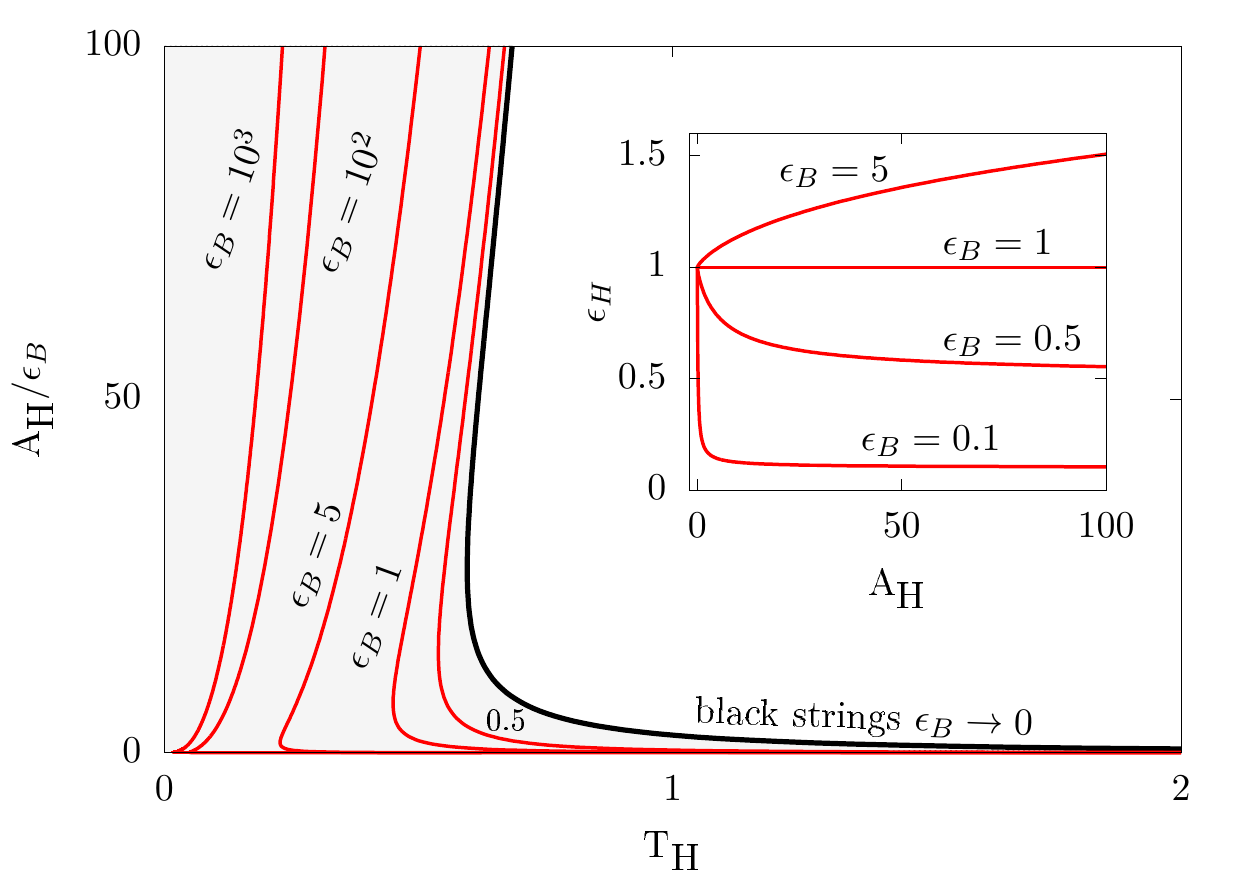}
}
\end{center}
\caption{The $(T_H,A_H)$-domain of existence of static vacuum black holes
with a deformed sphere at infinity.
The inlet shows the horizon deformation $\epsilon_H$ for several values of the boundary squashing $\epsilon_B$. 
}
\label{fig1}
\end{figure}

The BHs are  obtained numerically,
by solving the field equations subject to the
boundary conditions described above \cite{numerics},
the results being displayed in units with $L=1$.
All configurations reported here
(including the SUSY ones)
 are regular on and outside the horizon.
Also, they do not present other pathologies (such as closed timelike curves (CTCs)).
 
The most remarkable feature of these BHs is that they 
interpolate between two classes of solutions
with different topologies:
black strings and black branes.
Starting 
with the vacuum static case \cite{static}, we exhibit 
in Fig.\ref{fig1} the domain of existence of the BHs in the $(A_H,T_H)$-plane,
which shows that the  $\epsilon_B=1$ pattern is generic.
Their horizon deformation is also displayed,
and one can see that the parameters $\epsilon_H$ and $\epsilon_B$ are correlated,
with  $\epsilon_H/\epsilon_B \to 1 $ for large horizon size
{(for instance this happens in Fig.\ref{fig1} also in the $\epsilon_B=5$ curve for very large values of the area, which are not displayed in the range shown). }
However, the small BHs are always close to sphericity, 
with a well-defined  $L_H\to 0$ limit.
This is a smooth, horizonless geometry,
which can be viewed as a deformation of the globally AdS spacetime,
providing a natural $background$ for a model with $\epsilon_B\neq 1$.
Moreover, the ratio $\epsilon_H/\epsilon_B$  is well defined as
$\epsilon_B\to 0$,
 depending only on the value of $L_H$.
Then, after the rescaling $\psi\to \bar \psi/\epsilon_B$,
one finds that  in the $\epsilon_B\to 0$ limit, 
the solutions become 
AdS {\it black strings}.
For such configurations,
both the boundary and the horizon metric
are the direct product $S^2\times S^1$,
with an arbitrary periodicity for $S^1$
as parametrized by $\bar \psi$ 
($e.g.$
 $
ds^2_{{\cal B}}=\frac{L^2}{4}
\left(
d\theta^2+\sin^2\theta d\phi^2+d\bar \psi^2
\right)).
$
These solutions were originally found in 
\cite{Copsey:2006br}
(see also \cite{Mann:2006yi},
\cite{Bernamonti:2007bu}),
and provide natural AlAdS generalizations of the 
(uniform) 
 black strings in  $d=5$
Kaluza-Klein theory.
We also notice that $\epsilon_B^2$ can be continued to negative
values, which results in a different set of solutions 
with CTCs.

The infinitely squashed limit 
is also well defined.
After 
taking
$\epsilon_B \to \lambda N$
together with a
rescaling  the coordinates,
$r \to \lambda r$, 
$\theta \to \Theta/\lambda  $,
$\psi \to  -\Psi/(N\lambda^2) -\phi$,
$t \to t/ \lambda$, 
one finds that $\lambda \to \infty$
results in 
an AlAdS 'twisted' {\it black brane} \cite{DHoker:2009ixq}, with a 
conformal boundary metric which is the product of time and 
\begin{eqnarray}
\label{planarB}
ds^2_{{\cal B}}=\frac{L^2}{4}\left(d\Theta^2+\Theta^2 d\phi^2+(d\Psi+N \frac{\Theta^2}{2} d\phi )^2 \right).
\end{eqnarray} 
The same type of line-element is found for the horizon metric 
(although with different factors for the two distinct parts): 
\begin{align}
\label{planarH}
ds^2_{{\cal H}}=\frac{L_H^2}{4}\left(d\Theta^2+\Theta^2 d\phi^2+m_H(d\Psi+N \frac{\Theta^2}{2} d\phi )^2 \right)~.
\end{align} 

\begin{figure}[t!]
\begin{center}
\mbox{\hspace*{0.cm}
\includegraphics[height=.25\textheight, angle =0]{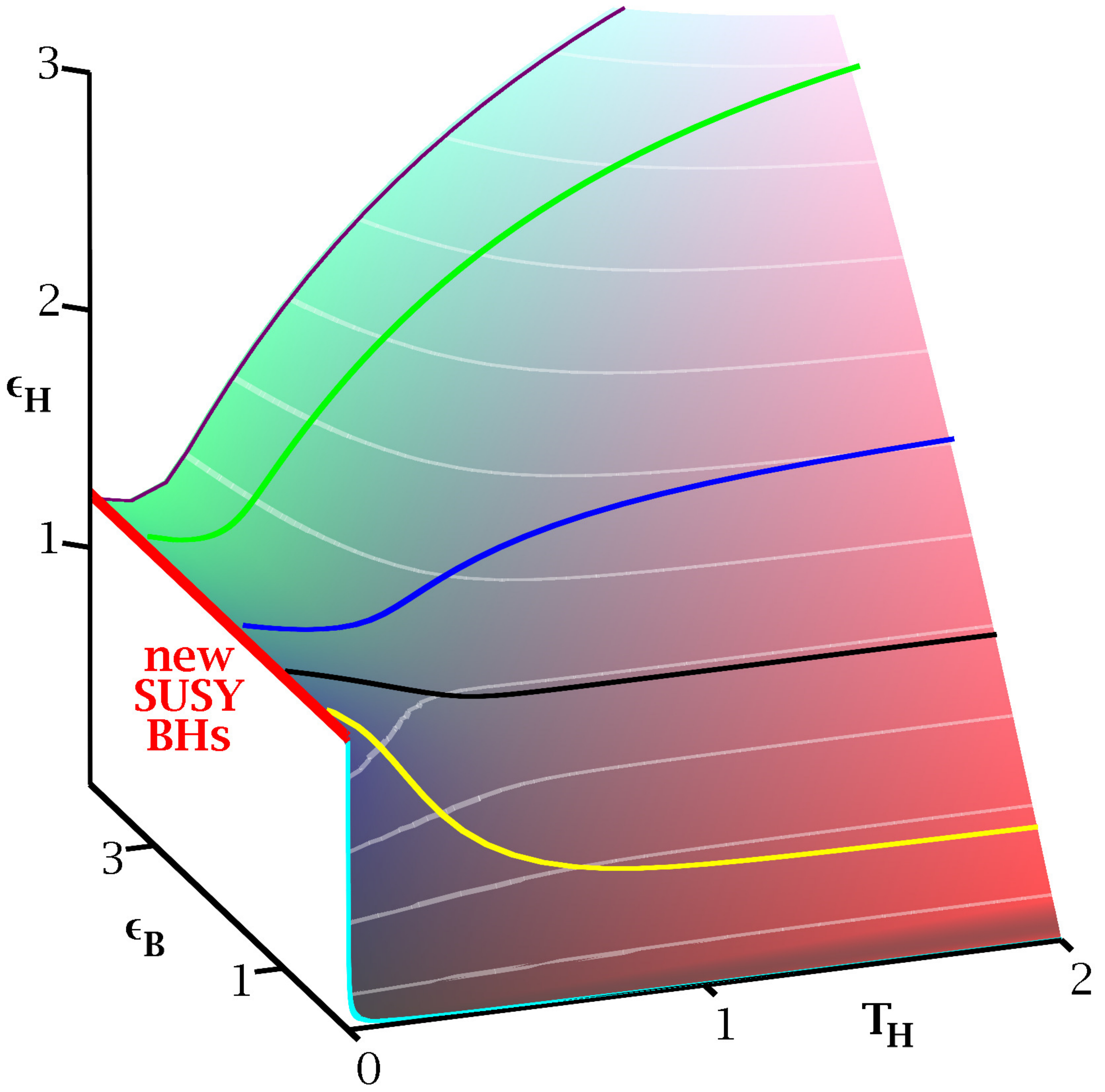}
}
\end{center}
\caption{ The $(T_H,\epsilon_B,\epsilon_H)$-domain of existence is shown for a 
particular set of black holes.
The $(Q,J,\Phi_m)$-dependence  
 of the boundary squashing parameter $\epsilon_B$
is imposed
such that the extremal limit corresponds to supersymmetric solutions
($e.g.$ $\Phi_m=L(\epsilon_B^2-1)/2\sqrt{3}$).   
Color curves are isolines of constant $\epsilon_B$, and grey curves are isolines of constant $\epsilon_H$. 
}
\label{fig2}
\end{figure}

\begin{figure}[t!]
\begin{center}
\mbox{\hspace*{0.cm}
\includegraphics[height=.25\textheight, angle =0]{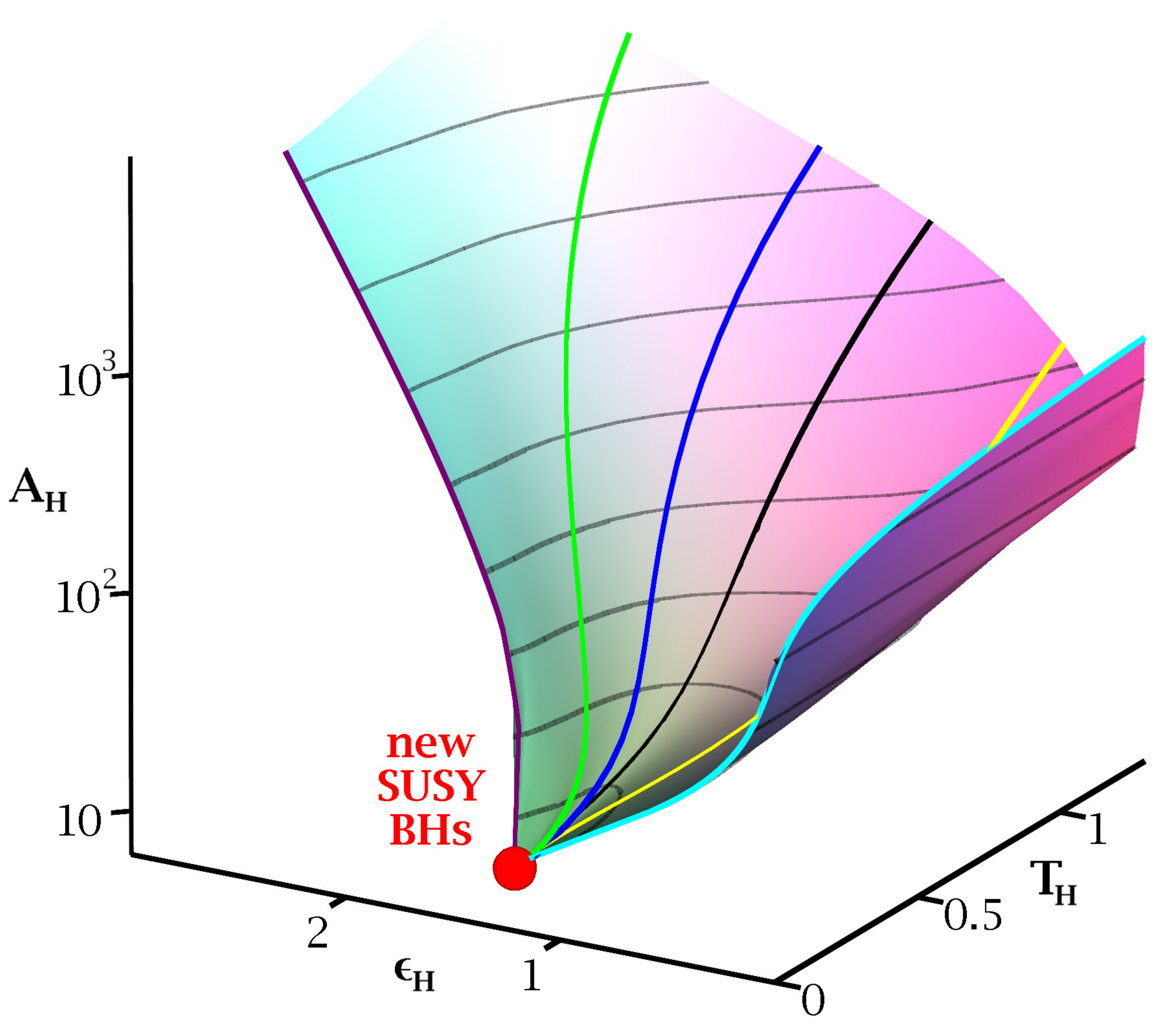}
}
\end{center}
\caption{The $(T_H,\epsilon_H,A_H)$-diagram for the same solutions as in Fig. \ref{fig2}.
The full set of SUSY solutions is mapped here to a single point, possessing the same 
horizon area and deformation. 
Isolines of constant $\epsilon_B$ and $A_H$ are marked with color and grey lines, 
respectively. 
}
\label{fig3}
\end{figure}

In the generic spinning magnetized case, 
the relation 
 between the horizon and boundary deformations
 is more intricate.
Depending on the values of $M,J,Q$ and $\Phi_m$, 
 one finds  $e.g.$
solutions with a large $\epsilon_B$
and arbitrarily small $\epsilon_H$.
Similarly, there are  spinning magnetized BHs
whose horizon is a round sphere, $\epsilon_H=1$,
while the value of $\epsilon_B$ is very large (or very small).
Some of these features
can be seen in Fig.\ref{fig2}, 
where the  ($\epsilon_B,\epsilon_H)$-diagram 
is shown vs. $T_H$
for a particular set of solutions.
In this Figure it can be seen that the black holes reach a finite value of $\epsilon_H$ as we approach extremality $(T_H \to 0)$. This is always the case as long as $\epsilon_B$ is different from zero. On the other hand, in this figure we can see that for non-extremal black holes (with $T_H\neq 0$),  decreasing the value of  $\epsilon_B$ to zero also makes $\epsilon_H $ goes to zero. This indicates the presence of a regular string limit, with $\epsilon_H/\epsilon_B$ finite
(note $e.g.$
the linear relation between 
$\epsilon_B$ and $\epsilon_H$ for large enough values of the temperature).
In Fig. \ref{fig3} we show the area $A_H$ as a function of $\epsilon_H$ and $T_H$.
In this Figure we can see that the $\epsilon_H \to 0$ limit of non-extremal solutions causes the area to go up to infinity (see how the surface bends up on the right side of the figure). Actually this limit has a finite $A_H/\epsilon_B$ limit, indicating that the density $A_H/\epsilon_B$ of the limiting string is finite. We have verified numerically these features by directly constructing the string configurations and comparing the corresponding charge densities (such as $M/\epsilon_B$, etc). 

Therefore the black string limit is well defined for 
a part of the parameter space only  
(for example, close to extremality 
the correlation between $\epsilon_H$ and $\epsilon_B$ is lost,
without a smooth black string limit in the extremal case).   
Nevertheless, the 
infinite squashing limit is well behaved, 
resulting in a family of charged and magnetized black branes 
with the same conformal boundary metric 
as in the static vacuum case \cite{Blazquez-Salcedo:2017ghg}. 
 
Let us mention  that, 
unsurprisingly,
the squashed spinning and magnetized BHs share some common features with the
  $\epsilon_B=1$ solutions in 
\cite{Cvetic:2004hs},
\cite{Blazquez-Salcedo:2017cqm}.
For example,  
their thermodynamics is qualitatively similar to that case,
the BHs with a large enough boundary magnetic field
becoming thermodynamically stable for the full range of $T_H$, see Fig. \ref{fig3}.
Also, for any
$\epsilon_B>0$,
the zero horizon size limit of the solutions with $\Phi_m\neq 0$ is nontrivial and 
describes a one parameter family of spinning charged (non-topological) solitons.
Such solutions possess no horizon, 
with the size of both
parts of the  horizon metric
vanishing as $r_H\to 0$
(while $g_{rr}$ and $g_{tt}$ remain nonzero).
Interestingly,
for a given $\epsilon_B$,
the solitons form a one parameter family of solutions, most of their properties
being determined by $\Phi_m$.
For example, 
the following relation  holds
\begin{eqnarray}
 J=  \Phi_m Q,~~{\rm with}~~Q=-\frac{8\pi \Phi_m^2}{3\sqrt{3}},
\end{eqnarray}
as implied by the existence of two first integrals of the system  \cite{Blazquez-Salcedo:2017ghg}.
Although no similar expression exists for $M$, 
the relation
\begin{eqnarray}
\label{mass-fit}
M= \frac{3\pi L^2}{32}\left(
1+\frac{32}{\sqrt{3}}\frac{\Phi_m}{L}(\epsilon_B^2-1)-\frac{64}{3}\frac{\Phi_m^2}{L^2}\epsilon_B^2
\right),
\end{eqnarray}
provides a good fit for the mass of the solutions
  with $\epsilon_B$ close to one and a small boundary magnetic field.

{\bf Supersymmetric black holes.--}
As found in  Ref. \cite{Gutowski:2004ez},
a particular set of $\epsilon_B=1$ BHs
	\cite{Cvetic:2004hs}  
	preserves one quarter of the supersymmetry.
Then it is natural to inquire 
if these special configurations survive when deforming the boundary geometry 
according to (\ref{boundary-sphere}).
To address this issue, we use the framework proposed in 
 \cite{Gutowski:2004ez}
and consider a line element 
\begin{eqnarray}
\label{metric0}
&&ds^2=-f^2(\rho) \left( dt+\Psi(\rho)  {\ \sigma}_3 \right)^2
+\frac{1}{f(\rho)} ds_B^2~,
\\
\nonumber
&&
{\rm ~~~~with}~ds_B^2=
   d\rho^2+a^2(\rho)( { \sigma}_1^2+  {\sigma}_2^2)+ b^2(\rho)  {\sigma}_3^2~,
\end{eqnarray}
and a U(1) potential 
\begin{eqnarray} 
\label{A-susy}
A= \frac{\sqrt{3}}{2} \left[ f(\rho) dt + \left(f(\rho)\Psi(\rho)+\frac{L}{3}p(\rho) \right)  {\sigma}_3  \right ].
\end{eqnarray}
All functions which enter the above Ansatz
are determined by $a(\rho)$ and its derivatives \cite{Gutowski:2004ez},
$a(\rho)$ being the solution of a sixth order equation 
\begin{eqnarray} 
\label{eqa}
\Big(\nabla^2 f^{-1} +8 L^{-2}f^{-2} - \frac{L^2 g^2}{18} + f^{-1} g \Big)' + \frac{4a'g}{af} = 0,~{~~~} 
\end{eqnarray}
where
$\nabla^2 $
is the Laplacian for $ds_B^2$,  
$
 g =-\frac{a'''}{a'} -  \frac{3a''}{a} - \frac{1}{a^2} + \frac{(4a')^2}{a^2} ,
$
and
$
f^{-1} = \frac{L^2}{12 a^2 a'}(4 (a')^3 + 7 a\, a' a'' - a' + a^2 a''').
$
Any solution of this equation  
 corresponds to a 
 configuration 
which preserves (at least) one quarter of the supersymmetry.

Without any loss of generality,
the horizon is located at $\rho=0$,
with a Taylor series expansion of the solution
$
a(\rho)= L \sum_{k\geq 1} \alpha_k (\frac{\rho}{L})^k.
$
Combining this expansion with the sixth order eq. (\ref{eqa}), 
one finds 
 $\alpha_1\neq 0$,
$\alpha_2 =0$,
together with the constraint
\begin{eqnarray}
\label{nh-susy-r2}
 (11 \alpha_1^2-8) \alpha_4 =0.
\end{eqnarray}
The choice $\alpha_4=0$ corresponds to the exact solution
found by  Gutowski and Reall  
 in Ref.\cite{Gutowski:2004ez},
with
$a(\rho)=\alpha L \sinh(\rho/L)$,
where $\alpha_1=\alpha>1/2$. 
These BHs possess a horizon with 
$L_H=L\sqrt{(4\alpha^2-1)/3}$
and $\epsilon_H=\sqrt{\alpha^2+3/4}>1$,
while $\epsilon_B=1$ and $\Phi_m=0$.

However, the condition  (\ref{nh-susy-r2})
can also be satisfied by taking 
$\alpha_1=2\sqrt{2/11}$, $\alpha_4\neq 0$. 
This
leads to a new set of BHs
with the following near-horizon
expansion \cite{CM}
\begin{eqnarray}
\label{sola0n}  
\frac{a(\rho)}{L}= 2\sqrt{\frac{2}{11}}\frac{\rho}{L}
+ \alpha_3 (\frac{\rho}{L})^3
+ \alpha_4 (\frac{\rho}{L})^4
+\dots~.
\end{eqnarray} 
These asymptotics can be 
smoothly matched to  a
large-$\rho$ expansion 
of $a(\rho)$
with the following leading order terms
\begin{eqnarray} 
\label{inf-rho} 
&&
\frac{a(\rho)}{L}= 
 a_0 e^{\frac{\rho}{L}} + (  a_2 +  c \frac{\rho}{L})
\frac{e^{-\frac{\rho}{L}}}{  a_0}   
\\
\nonumber
&&
+\left(  a_4 
+ \frac{2- 16a_2 -5  c}{12}c\frac{\rho}{L} 
-\frac{2}{3} c^2 (\frac{\rho}{L})^2\right) \frac{e^{-\frac{3\rho}{L}}}{a_0^3} 
+\dots~,
\end{eqnarray} 
$
\{\alpha_3,\alpha_4;a_0,a_2,a_4 \}
$
%
in the above relations
being free parameters 
(with $a_0 \neq 0$) and $c=(1-\epsilon_B^2)/4$.

\begin{figure}[t!]
\begin{center}
\mbox{\hspace*{0.cm}
\includegraphics[height=.25\textheight, angle =0]{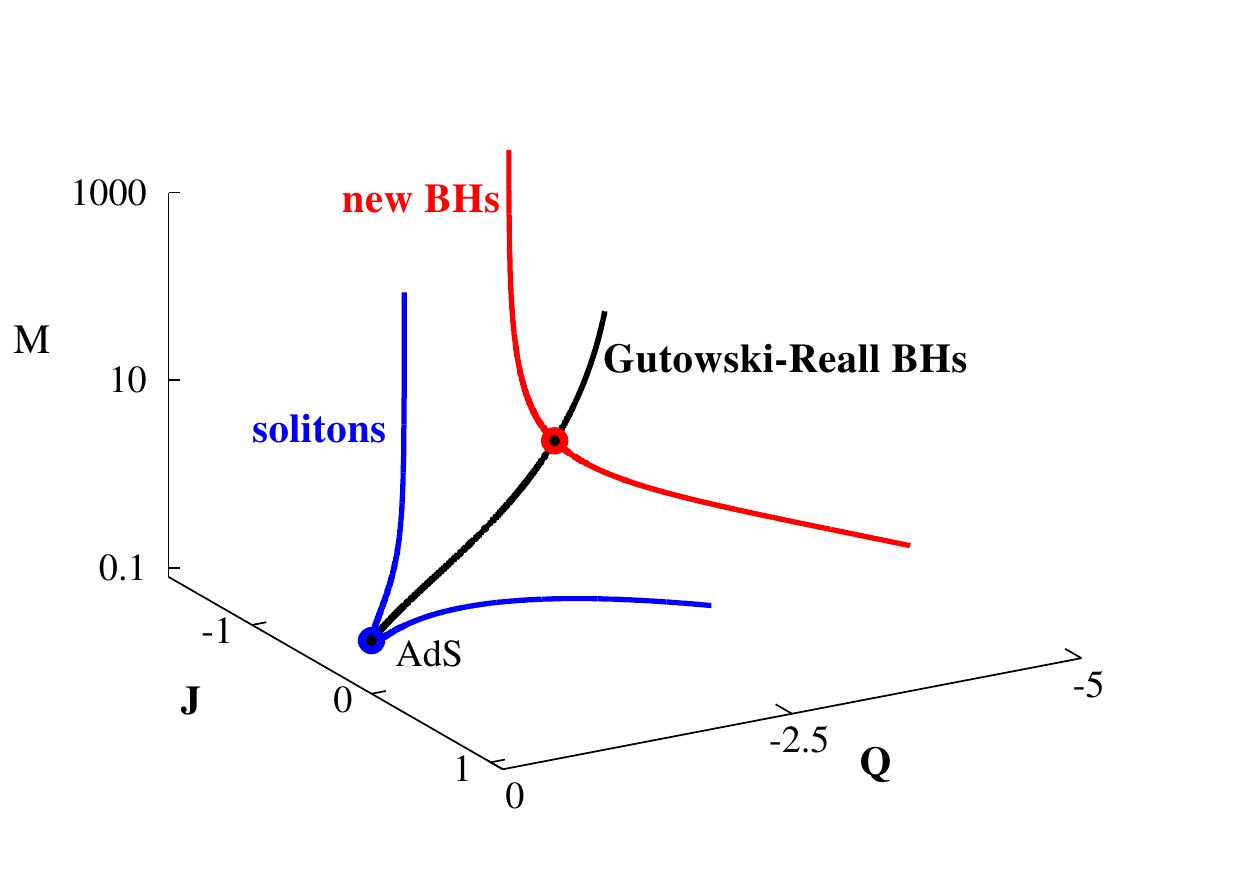}
}
\end{center}
\caption{The $(Q,J,M)$-diagram 
 is shown for three families of supersymmetric AdS solutions.
}
\label{fig4}
\end{figure}
This results in a family of AlAdS BHs,
which,
after moving to a non-rotating frame at infinity,
can also be viewed as a special class within the Ansatz
(\ref{metrici}).
Their spatial boundary is a squashed 
 sphere,
with $F_3/F_2\to \epsilon_B^2$.
The global charges are
determined by the
squashing parameter $\epsilon_B$, with 
\cite{Blazquez-Salcedo:2017ghg}
\begin{eqnarray}
&&
\nonumber
M=\pi L^2
\left(
\frac{7913}{34848}+\frac{33280}{35937}\frac{1}{\epsilon_B^2}-\frac{7}{36}\epsilon_B^2+\frac{89}{864}\epsilon_B^4
\right),
 \\
&&
\label{MJQ-susy-BH}
J=-\pi L^3
\left(
 \frac{16640}{35937}
-\frac{2795}{8712}\epsilon_B^2
+\frac{1}{9} \epsilon_B^4
-\frac{1}{27}\epsilon_B^6
\right),
\\
&&
\nonumber
Q= -\pi \sqrt{3} L^2
\frac{1}{13068}
 \left( 6449-1936\,{\epsilon_B}^{2}  + 968\,{\epsilon_B}^{4} \right) .
\end{eqnarray}
They necessarily possess a boundary magnetic field, with
$
 \Phi_m=\frac{2}{\sqrt{3}}(\epsilon_B^2-1),
$
while the
horizon angular velocity is
$\Omega_H=2/(L\epsilon_B^2)$.
The most unusual feature of the new BHs is  that 
although
$\epsilon_B$ is arbitrary, 
their
horizon geometry (\ref{horizon-metric})
 is 'frozen', 
with 
 $L_H= L \sqrt{7/11}$,
 $\epsilon_H=\sqrt{65/44}$ 
and 
\begin{eqnarray}
\label{AH-susy} 
A_{H}=7 \pi^2 L^3 \frac { \sqrt {455}}{121}~.
\end{eqnarray} 
As $\epsilon_B \to 1$,
the solutions
 bifurcate  from a critical  
Gutowski-Reall BH with
$\alpha=2\sqrt{2/11}$.
Also, as seen in Figs. \ref{fig2}, \ref{fig3}, 
they are  approached smoothly
as a particular limit of the general solutions.
However, different from the non-extremal case,
these BHs do not possess a solitonic limit.
Nevertheless, SUSY solitons
with $\epsilon_B\neq 1$ 
 exist as well
 \cite{Cassani:2014zwa}, 
satisfying a different set of boundary conditions at $\rho=0$
and bifurcating from the globally AdS background.
Again, most of their 
physical properties
are determined by the boundary squashing parameter $\epsilon_B$.
A diagram summarizing the picture for 
these three different types of SUSY solutions 
is shown in Fig. \ref{fig4}.

The 'frozen' horizon geometry prevents the SUSY solutions to approach
a black string limit as $\epsilon_B\to 0$. 
Instead, a BH with non-asymptotically flat,
non-AlAdS asymptotics is approached.
The limit $\epsilon_B\to \infty$ is also non-standard.
The same scaling as in the
non-SUSY case 
 leads to an exact plane-fronted wave solution
which is not asymptotically AdS
 \cite{planar}.

{\bf Further remarks.--}
%
The new 
rotating magnetized BH solutions in this work provide new backgrounds
whose AdS/CFT duals describe four-dimensional field
theories in a squashed Einstein universe.
Also, they  can
be uplifted either to type IIB or to eleven-dimensional supergravity
\cite{Chamblin:1999tk},
\cite{Gauntlett:2004zh},
\cite{Gauntlett:2006ai},
\cite{Gauntlett:2007ma}.
 
Their existence raises many questions.
In particular, it would be interesting to provide a microscopic interpretation
from the boundary CFT for the entropy
of the SUSY BHs.
Generalizations 
of these solutions 
with an arbitrary multipolar structure
of the U(1) field at infinity 
and two independent rotation parameters
are also likely to exist, in particular SUSY configurations.



 \medskip
{\bf Acknowledgement}
We gratefully acknowledge support by the DFG, 
in particular, the DFG Research Training Group 1620 ''Models of Gravity''. 
E. R. acknowledges funding from the FCT-IF programme and is 
also partially supported by  the  H2020-MSCA-RISE-2015 Grant No.  StronGrHEP-690904, 
and by the CIDMA project UID/MAT/04106/2013. F. N.-L. acknowledges funding from Complutense University - Santander under project PR26/16-20312.



\begin{thebibliography}{000}
\bibitem{gauge}	
	The   gauge freedom in the line-element (\ref{metrici})  
is fixed by taking
	$F_2/F_1=r^2(1+r^2/L^2)$.
\bibitem{numerics}	
 We employ a collocation method for boundary-value ordinary
differential equations, equipped with an adaptive
mesh selection procedure. Typical mesh sizes include
 around $10^4$, the typical relative accuracy of the solutions being
 around $10^{-10}$.
\end{thebibliography}
\end{document}